\documentclass[12pt]{article}
\usepackage[scaled]{helvet}
\usepackage[margin=1in]{geometry}
\usepackage[nodisplayskipstretch]{setspace}

\usepackage[utf8]{inputenc}
\usepackage{amsmath}
\usepackage{amsfonts}
\usepackage{multicol, booktabs, bbm, color, xcolor}
\usepackage{natbib}
\usepackage{graphicx}
\usepackage{subfigure}
\graphicspath{{./img/}}
\usepackage{titlesec}
\titleformat{\section}{\large\bfseries}{\thesection}{1ex}{}
\titleformat{\subsection}{\normalsize\bfseries}{\thesubsection}{1ex}{}
\titleformat{\subsubsection}{\normalsize\bfseries}{\thesubsubsection}{1ex}{}
\titlespacing{\section}{0em}{1em}{0em}
\titlespacing{\subsection}{0em}{1em}{0em}
\titlespacing{\subsubsection}{0em}{1em}{0em}

\title{Estimating Perinatal Critical Windows of Susceptibility to Environmental Mixtures
    via Structured Bayesian Regression Tree-Pairs}
\author{Daniel Mork \& Ander Wilson}
\date{2021}

\begin{document}
    \maketitle
    \setstretch{1.2}
    \noindent\textbf{Summary}
    
    \noindent Maternal exposure to environmental chemicals during pregnancy can alter birth and children’s health outcomes. Research seeks to identify critical windows, time periods when the exposures can change future health outcomes, and estimate the exposure-response relationship. Existing statistical approaches focus on estimation of the association between maternal exposure to a single environmental chemical observed at high-temporal resolution, such as weekly throughout pregnancy, and children’s health outcomes. Extending to multiple chemicals observed at high temporal resolution poses a dimensionality problem and statistical methods are lacking. We propose a tree-based model for mixtures of exposures that are observed at high temporal resolution. The proposed approach uses an additive ensemble of structured tree-pairs that define structured main effects and interactions between time-resolved predictors and variable selection to select out of the model predictors not correlated with the outcome. We apply our method in a simulation and the analysis of the relationship between five exposures measured weekly throughout pregnancy and resulting birth weight in a Denver, Colorado birth cohort. We identified critical windows during which fine particulate matter, sulfur dioxide, and temperature are negatively associated with birth weight and an interaction between fine particulate matter and temperature.  Software is made available in the \texttt{R} package \texttt{dlmtree}.
    
    
    \newpage
    \section{Introduction}\label{sec:intro}
    
   Maternal exposure to environmental chemicals during pregnancy is an important public health concern due to the potential impact on children’s health. Increased exposure to environmental chemicals has been linked to decreased birth weight, increased risk of asthma, and altered neurological development, among other outcomes \citep{Bosetti2010AmbientLiterature, Stieb2012AmbientMeta-analysis, Jacobs2017TheReview}. There is also evidence that changes in temperature are associated with  birth weight \citep{Jakpor2020TermVariability, Kloog2015UsingMassachusetts}. Recent research has focused on leveraging exposure data that is observed at high temporal resolution throughout pregnancy to identify critical windows of susceptibility during the gestational process \citep{Wright2017EnvironmentHealth}.   Critical windows are periods in time when an exposure can alter a future health outcome and could be as short as a week or could span many months. Estimating the exposure-response relationship over the course of pregnancy is an equally important goal. Most studies that leverage repeated measurements of exposure  consider the effects of only a single environmental chemical. While understanding the effects of a single exposure is essential, considering a mixture is necessary to develop a more realistic picture of the exposure-response relationship because it accounts for interactions and controls for confounding by co-exposures \citep{Davalos2017CurrentExposures}. In this paper, \textit{mixture} refers to simultaneous exposure to multiple environmental factors. As an example of the importance of considering interactions and mixtures, \citet{Anenberg2020SynergisticEvidence} documented evidence of a synergistic effect of heat and air pollution on cardiovascular and
   respiratory disease among adults. However, estimating critical windows for mixtures has been elusive due to an absence of statistical methods.
   
   A distributed lag model (DLM) is commonly used to estimate the association between a single time-resolved exposure and a health outcome \citep{ Zanobetti2000, Warren2012Spatial-TemporalExposure, Wilson2017a, Gasparrini2017}. A DLM regresses the outcome on the exposure measurements at multiple time points, e.g. regressing birth weight on weekly mean exposure during gestation. Because of high temporal correlation between repeated measures of exposure there is a need to regularize the DLM. This is most commonly done with a constrained DLM that smooths the effect over the exposure time and allows for sharing of information between parameters at different times during gestation. Common constraints include splines \citep{Zanobetti2000, Gasparrini2017}, Gaussian processes \citep{Warren2012Spatial-TemporalExposure} and principal components \citep{Wilson2017a}. Compared to using average exposure over pregnancy or each of the trimesters, the DLM has been shown to reduce bias in estimates as well as improve critical window estimation \citep{Wilson2017PotentialHealth}. To account for multiple exposures, the DLM can be used additively and extended to include interactions between two time-resolved predictors \citep{Chen2019DistributedPollutants, Muggeo2007BivariateMortality}. However, methods to identify critical windows and estimate exposure-time-response functions with mixtures of more than two time-resolved predictors are lacking.
   
   Estimating DLMs with interactions requires flexibility to identify interactions between exposures across time. Interactions at different time points correspond to the `priming' hypothesis. Priming posits that prenatal exposure leads to phenotype changes \citep{Bolton2014PrenatalOffspring}. These changes may result in altered biological mechanisms that increase susceptibility to later exposures. Extending DLMs to include time-sensitive interactions poses a challenge of increasing dimensionality in the required parameter space. As the number of exposures and temporal resolution of measurements increase, the number of possible interactions increases at a quadratic rate. In addition, data for mixtures observed at multiple time points typically exhibit high collinearity both over repeated measures and across mixture components at a single time point. A key challenge is, therefore, regularizing the model while allowing for flexibility to identify critical windows.
   
   We propose a regression tree approach to estimate a constrained DLM for a single exposure or mixture of exposures observed at multiple time points. Regression trees have been applied in numerous fields including the study of chemical mixtures observed at a single time point in environmental epidemiology \citep{Park2017ConstructionNHANES}. Bayesian additive regression trees (BART), introduced by \citet{Chipman2012}, has been adapted for a wide variety of data generating situations such as high dimensional prediction \citep{Linero2018} or causal inference \citep{Hahn2020BayesianEffects}. While BART is generally focused on out-of-sample prediction, our goal is to adapt this framework to the estimation of distributed lag effects. Applied to exposure observations taken repeatedly over time, current regression tree techniques are lacking in several respects. First, they would treat the measurements for a single exposure at adjacent times as independent predictors. This is equivalent to fitting an unconstrained DLM which is unstable due to high collinearity. Second, regression trees would not account for the structure in mixture data where one measurement from each exposure is taken at the same time point. To account for temporal ordering in a single time-resolved predictor, \citet{Mork2021TreedModels} proposed a tree-based model to estimate a distributed lag nonlinear model that subdivides the time and exposure-concentration dimensions of the exposure-time response surface, but fails to generalize to mixtures.
   
   In this paper, we define the distributed lag mixture model (DLMM), which extends the DLM to estimate the main effects of multiple exposures along with all two-way interactions. We propose regression tree methods for estimating both DLM and DLMM. Our DLM method uses regression trees that split on time to construct a constrained DLM that is piecewise constant with breakpoints determined by binary splits in the tree. Combined in an additive ensemble of trees, the resulting DLM can be approximately smooth or have piecewise constant distributed lag effects.  Our proposed DLMM further builds on the tree literature. We introduce the concept of tree pairs---two trees that collectively define DLM structures of main effects and interactions between two time-resolved exposure measurements. We develop a computational framework to estimate an additive ensemble of tree pairs that allows for both the tree structures and the exposures to which the tree structures are applied to be learned from the data. Furthermore, our method conducts exposure selection and effect shrinkage to remove time-resolved predictors or interactions that do not influence the outcome. 
   
   We apply our models to a Colorado-based administrative birth cohort. This analysis investigates changes to birth weight associated with five environmental exposures measured weekly during gestation. Software is made available in the \texttt{R} package \texttt{dlmtree}.

    \section{Colorado Birth Cohort Data}\label{sec:data}
    
    We analyze birth weight for gestational age $z$-score, BWGAZ, using birth vital statistics records from Colorado, USA. The data includes all births from Colorado with estimated conception dates between 2007 and 2015, inclusive. We limit the analysis to the Denver metropolitan area. Besides birth outcomes, the data includes individual covariates including mother's age, weight, height, income, education, marital status, prenatal care habits, smoking habits, as well as race and Hispanic designations. 
    
    We are interested in the association between birth weight and a mother's weekly exposure to particulate matter smaller than 2.5 microns in diameter (PM$_{2.5}$), nitrogen dioxide (NO$_2$), sulfur dioxide (SO$_2$), carbon monoxide (CO), and temperature. We limited our analysis to singleton, full-term births ($\geq37$ weeks) and observations with complete covariate and exposure data, resulting in 195,701 births. This study was approved by the Institutional Review Board of Colorado State University. Demographic breakdowns and additional data details are described in Web Appendix A.

    \section{Model}
    
    \subsection{Distributed lag mixture models}
    For a sample $i=1,\ldots,n$, let $y_i$ denote a continuous response, $\mathbf{x}_{i}=[x_{i1},\ldots,x_{iT}]'$ represent a vector of exposure measurements taken at equally spaced times $t\in\{1,\ldots,T\}$, and $\mathbf{z}_i$ represent a vector of covariates including model intercept. The single exposure DLM is
    \begin{equation}\label{eq:dlm}
        y_i=\sum_{t=1}^T x_{it}\theta_t+\mathbf{z}_i'\boldsymbol\gamma+\epsilon_i.
    \end{equation}
    In (\ref{eq:dlm}), $\theta_t$ is the linear effect of exposure at time $t$; $\boldsymbol\gamma$ is a vector of regression coefficients; and $\epsilon_i$ represents independent errors distributed $\mathcal{N}(0,\sigma^2)$.
    
    We consider a model involving $M$ exposures. Let $\mathbf{x}_{im}=[x_{im1},\ldots,x_{imT}]'$ represent the vector of measurements for exposure $m$ corresponding to individual $i$. A DLMM with pairwise interactions can be written
    \begin{equation}
        y_i = \sum_{m=1}^M\sum_{t=1}^T x_{imt}\theta_{mt}
        +\sum_{m_1=1}^M \sum_{m_2=m_1}^M \sum_{t_1=1}^T\sum_{t_2=1}^T x_{im_1t_1}x_{im_2t_2}\theta_{m_1m_2t_1t_2}
        +\mathbf{z}_i'\boldsymbol\gamma + \epsilon_i.
        \label{eq:main}
    \end{equation}
    Here, $\theta_{mt}$ is the main effect of exposure to pollutant $m$ at time $t$. Interactions are considered at every time combination $t_1$ for exposure $m_1$ with $t_2$ for exposure $m_2$ and parameterized by $\theta_{m_1m_2t_1t_2}$. This includes interactions within the same exposure. Within-exposure interactions at $t_1=t_2$ represents a quadratic main effect. In total, the DLMM requires $MT+{M+1\choose 2}T^2$ parameters, which quickly becomes a `large-$p$' problem as the number of exposures grows. For example, the DLMM in our data analysis involving 5 exposures and 37 time points requires estimating 20,720 parameters.
    
    \subsection{Treed DLM}\label{sec:tdlm}
\begin{figure}
        \centering
        \subfigure[]{
            \label{fig:tree-diagram-a}
            \includegraphics[height=5.5cm]{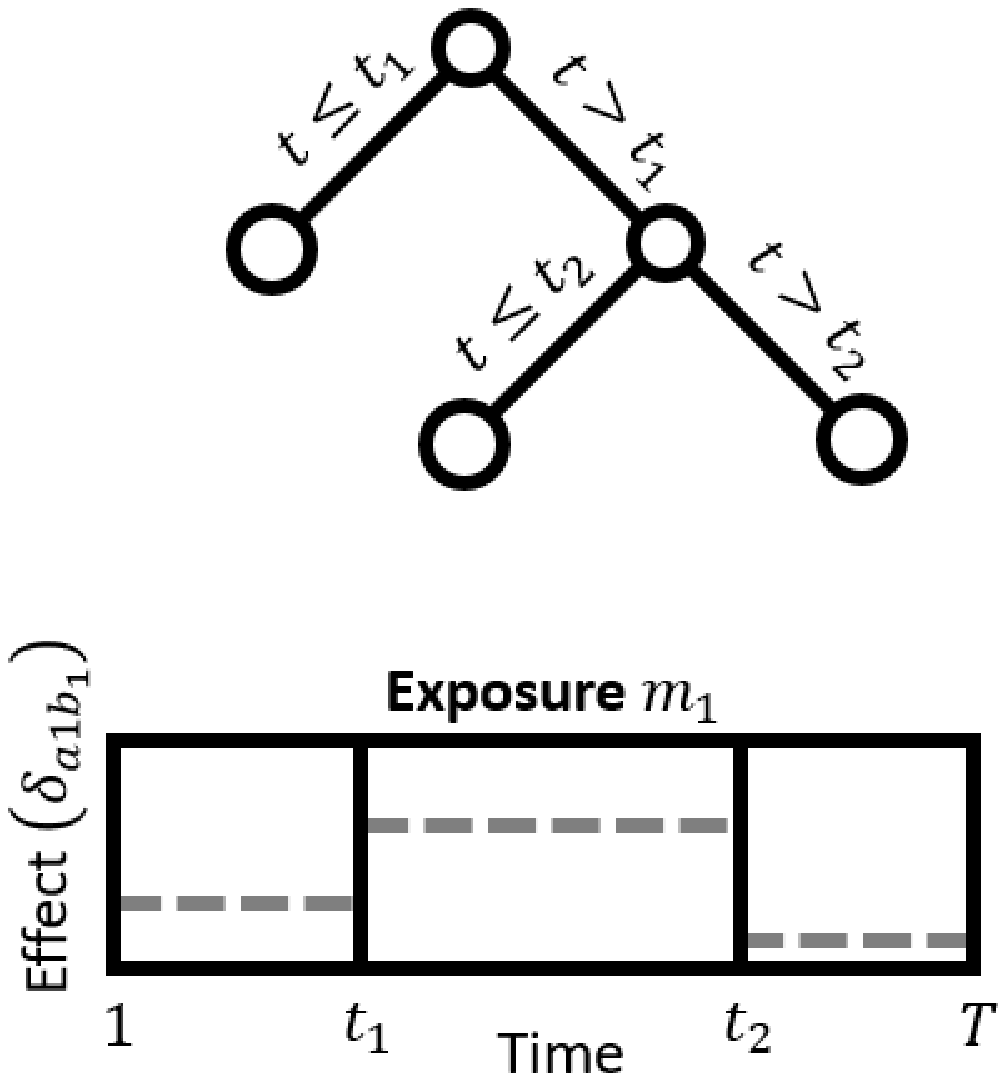}
        }
        \subfigure[]{
            \label{fig:tree-diagram-b}
            \includegraphics[height=5.5cm]{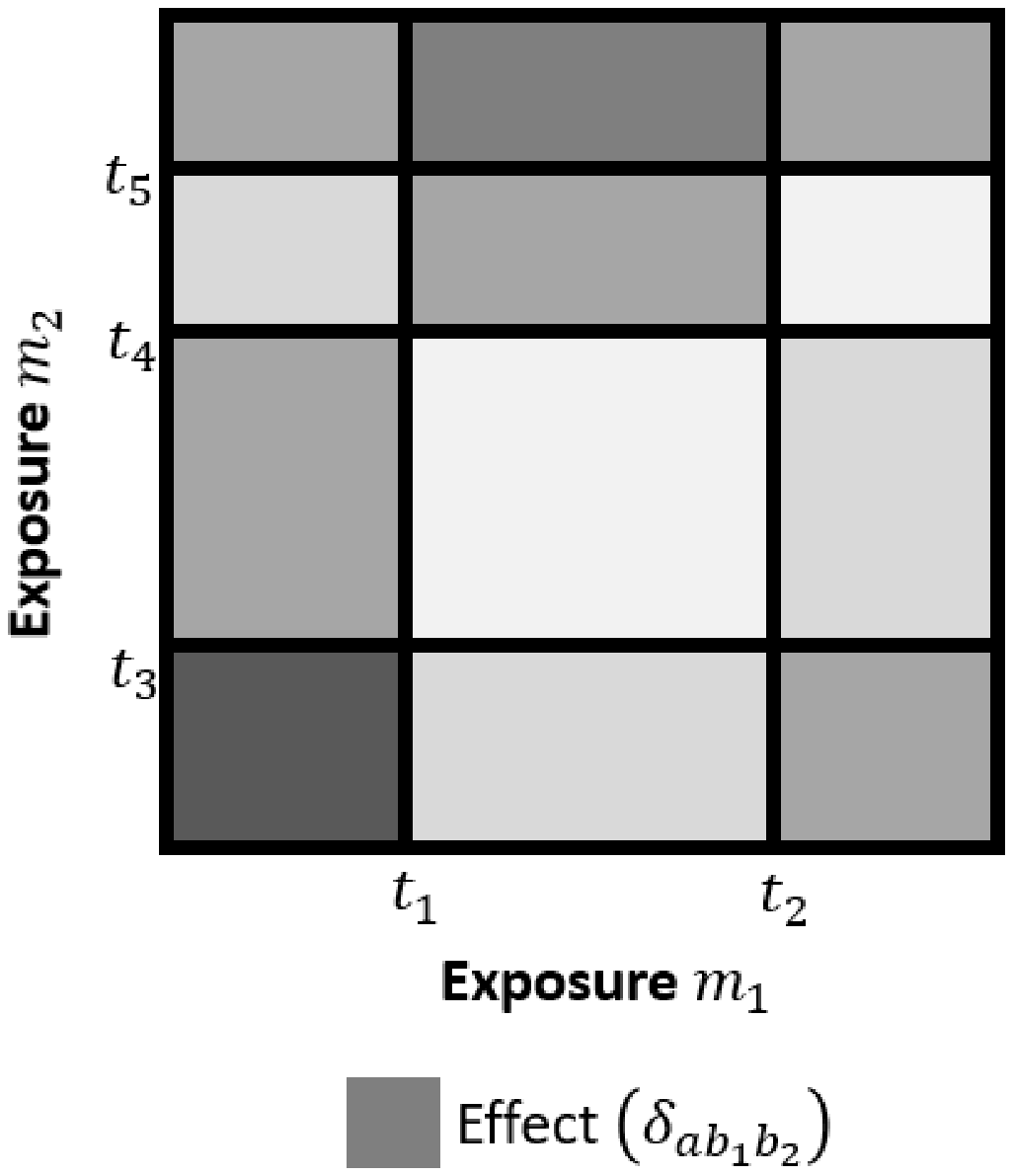}
        }
        \subfigure[]{
            \label{fig:tree-diagram-c}
            \includegraphics[height=5.5cm]{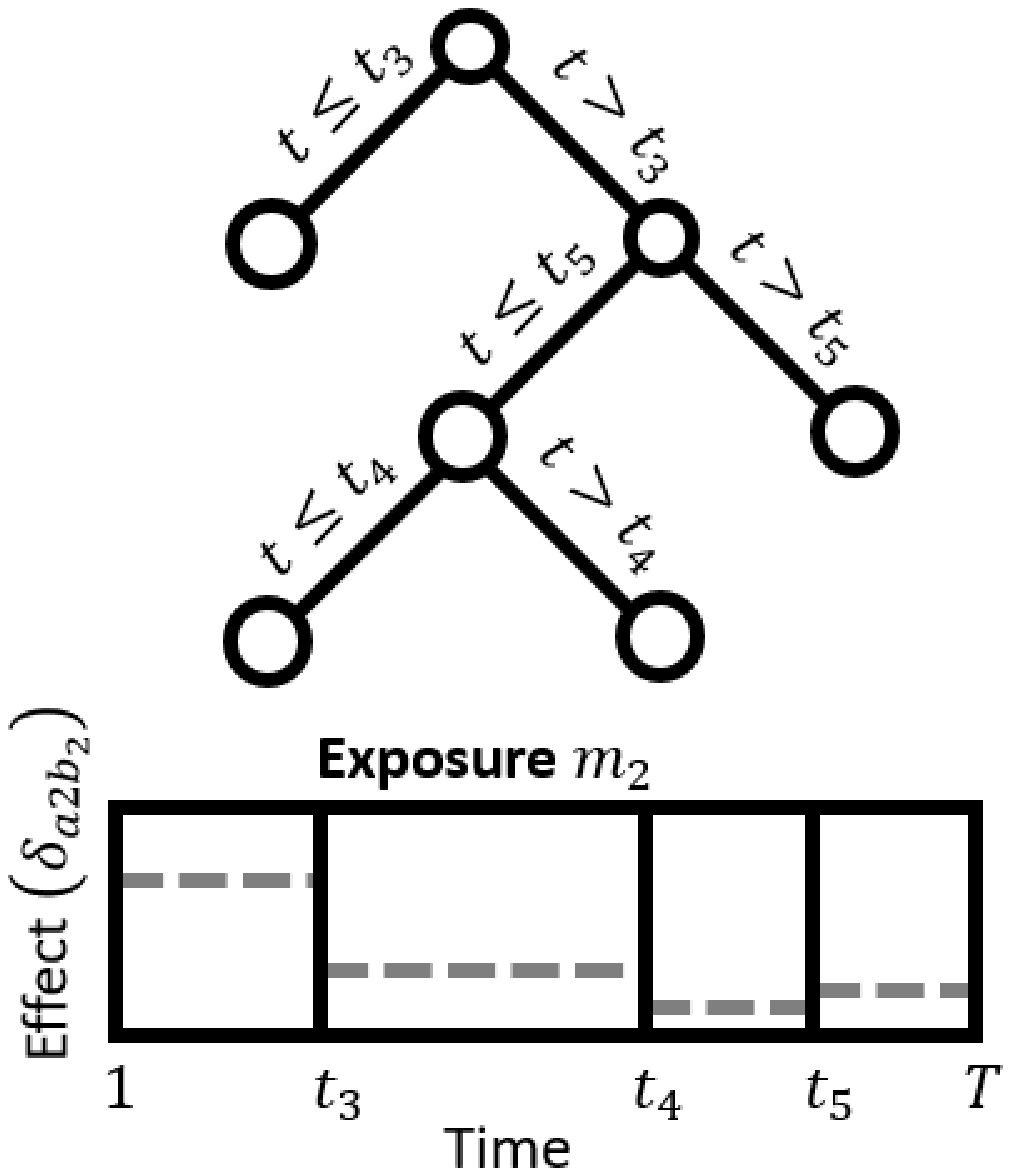}
        }
        \caption{Diagram of structured regression trees. Binary tree structures in panels (a) and (c) partition the time period for an individual exposure and each terminal node $\eta_{aib_j}$ corresponds to an estimated effect $\delta_{aib_j}$ (dashed lines). Panel (b) depicts how the structure of the two trees in a pair jointly define an interaction surface; the intersection of terminal nodes $\eta_{a1b_1}$ in panel (a) and $\eta_{a2b_2}$ in panel (c) has corresponding effect $\delta_{ab_1b_2}$ (with independent magnitude indicated by shading).}
        \label{fig:tree_diagram}
    \end{figure}

    We first introduce our proposed method of estimating a DLM for a single exposure with no interactions. In our treed distributed lag model (TDLM), binary trees partition the entire exposure time span, $T$, into non-overlapping segments. Figure \ref{fig:tree-diagram-a} illustrates the approach.  Each binary tree is characterized by a set of internal nodes that split on available time points and a set of terminal nodes, which are the endpoints of the tree structure. The terminal nodes define the time partition and are denoted $\eta_b$ for $b\in\{1,\ldots,B\}$. We define the linear relationship between each of the exposure measurements in terminal node $\eta_b$ and the outcome by a single coefficient, denoted $\delta_b$. The distributed lag effects in \eqref{eq:dlm} are, therefore, defined by the terminal nodes and node specific effects as $\theta_t=\delta_b\mbox{ if }t\in\eta_b$. This represents a piecewise constant DLM such that all times within the same terminal node have the same effect on the outcome. In our proposed model, both the tree structure and the effects are learned from the data.
    
    By keeping the number of terminal nodes small, TDLM introduces a necessary constraint on the distributed lag function to account for temporal correlation by assuming that the effect of exposures within the same terminal node are equal. This framework also allows for sharp changes in effect estimates across times at the boundary between different terminal nodes where the effect estimates can quickly change in magnitude or sign.  
    
    We propose an additive model with an ensemble of $A$ regression trees. This allows for smoothness in the DLM as each tree may partition the time span of exposure differently. For tree $\mathcal{T}_a$, $a\in\{1,\ldots,A\}$, with terminal nodes denoted $\{\eta_{ab}\}_{b=1}^{B_a}$ and corresponding effects $\mathcal{D}_a=\{\delta_{ab}\}_{b=1}^{B_a}$. The distributed lag effect $\theta_t=\sum_{a=1}^A g(t;\mathcal{T}_a,\mathcal{D}_a)$, where $g(t;\mathcal{T}_a,\mathcal{D}_a)=\delta_{ab}$ if $t\in\eta_{ab}$. This can be equivalently expressed as $\theta_t=\sum_{a=1}^A\sum_{b=1}^{B_a}\delta_{ab}\mathbb{I}(t\in\eta_{ab})$ where $\mathbb{I}(\cdot)$ is the indicator function.

    \subsection{Treed DLMM}
    The treed distributed lag mixture model (TDLMM) extends TDLM to multiple time-resolved predictors by replacing each individual tree with a pair of trees. A pair of trees is composed of two binary trees, which define the main effects of two exposures and the interaction between these same exposures. The interactions are structured based on the time partitions of each tree.  Exposures in each time-segment of the first tree are interacted with exposures in each time-segment of the second tree.  A diagram representing the structured regression trees and interaction surface is shown in Figure \ref{fig:tree_diagram}. The exposures partitioned by a tree pair may be two different exposures or the same exposure twice. In the case that both trees partition the same exposure, we estimate within-exposure interactions as well as potential nonlinear effects via quadratic terms created by same-time interactions. This maintains a hierarchy where an interaction is only included when the main effects are part of the model. Both the structures of the trees and exposure types used in the trees are learned in our model.
    
    In TDLMM we use an ensemble of $A$ tree pairs. The ensemble allows for different pairs of exposures to be included in the model, which correspond to the main effects and interactions that may be present. Consider tree-pair $\{\mathcal{T}_{a1},\mathcal{T}_{a2}\}$ with terminal nodes $\{\eta_{aib}\}_{b=1}^{B_{ai}}$ and corresponding effects $\mathcal{D}_{ai}=\{\delta_{aib}\}_{b=1}^{B_{ai}}$ for $i=1$ and $2$. Let $S_{ai}=m$ if exposure $m$ is partitioned by tree $\mathcal{T}_{ai}$. The main effect of exposure $m$ at time $t$ is
    \begin{equation}
        \theta_{mt}=\sum_{a=1}^A \sum_{i=1}^2 g(t;\mathcal{T}_{ai},\mathcal{D}_{ai})\mathbb{I}(S_{ai}=m)
    \end{equation}
    and the interaction between exposure $m_1$ and $m_2$ at times $t_1$ and $t_2$ is
    \begin{equation}
        \theta_{m_1m_2t_1t_2}=\sum_{a=1}^A
        g_I(t_1,t_2;\mathcal{T}_{a1}\times\mathcal{T}_{a2},\mathcal{I}_{a})\mathbb{I}(S_{a1}=m_1,S_{a2}=m_2).
    \end{equation}
    Here, $g_I(t_1,t_2;\mathcal{T}_{a1}\times\mathcal{T}_{a2},\mathcal{I}_{a})=\delta_{ab_1b_2}$ if $t_1\in\eta_{ab_1}$ and $t_2\in\eta_{ab_2}$, and $\mathcal{I}_{a}=\{\delta_{ab_1b_2}\}$ is the set of interaction effects corresponding to tree pair $a$.

    TDLMM can be reduced into two simpler models. The first drops within-exposure interaction by fixing these interactions to zero. The second simplification drops all interactions which is equivalent to an additive DLM. In the remainder of this paper we refer to these simplifications as TDLMMns (no-self interactions) and TDLMMadd (additive).

    \section{Prior Specification and Computation}
    
    \subsection{TDLM priors and posterior computation}\label{sec:TDLM-priors}
    The prior for TDLM consists of two parts: a prior on trees and a prior on the regression parameters conditional on the trees. We apply a normal prior to the regression parameters
    \begin{equation}\label{eq:prior-tdlm}
        \delta_{ab}|\tau_a^2,\nu^2,\sigma^2\sim\mathcal{N}(0,\tau_a^2\nu^2\sigma^2).
    \end{equation}
    Here, $\tau_a\sim\mathcal{C}^+(0,1)$ and $\nu\sim\mathcal{C}^+(0,1)$ define a horseshoe-like prior on tree-specific effects \citep{Carvalho2010}. This is a global-local shrinkage prior that performs shrinkage overall and at the tree level. The additional tree-specific variance component $\tau_a$ allows for the effects on poor fitting trees to be shrunk. When shrunk, the trees can more easily reconfigure before regaining larger terminal node effects. In practice, we find including this tree-specific variance results in more precise estimates of the distributed lag function. We specify global prior $\sigma\sim\mathcal{C}^+(0,1)$ and $\boldsymbol\gamma\sim{MVN}(\mathbf{0},\sigma^2c\mathbf{I})$, where $c$ is fixed at a large value.
    
    The prior on trees is a stochastic tree generating process based on the approach of \citet{Chipman1998}. Complete details are given in Web Appendix B.

    \subsection{TDLMM priors and posterior computation}\label{sec:tdlmm}
    For TDLMM with multiple predictors and tree pairs the prior involves three components: 1) the prior on trees; 2) the prior on node effects; and 3) the prior on which exposures appear in each tree pair. One goal of TDLMM is to shrink or remove exposure and interaction effects that are uncorrelated with the response. We specify a prior that uses two approaches to achieve this goal. First, we add a prior on node specific effects that allows for effects of unique exposures and interactions to be shrunk. Second, we specify a prior on which exposures are included into each of the $A$ tree pairs. This avoids having to pre-specify pairs of exposures into each tree pair and allows for variable selection because an exposure that is not included into any tree pair is selected out of the model.

    In TDLMM each of the trees in tree-pair $a$ is defined by a tree structure $\mathcal{T}_{ai}$, $i\in\{1,2\}$, and an exposure $S_{ai}$ that the tree structure is applied to. The prior on $\mathcal{T}_{ai}$ is the same as described in Section \ref{sec:TDLM-priors}. The prior distribution on exposure $S_{ai}$ is
    \begin{equation}
        \label{eq:exposure-prob}
        S_{ai}|\mathcal{E}
        \sim\mbox{Categorical}(\mathcal{E}),\quad
        \mathcal{E}\sim
        \mbox{Dirichlet}(\kappa,\ldots,
        \kappa).
    \end{equation}
    Here, $\mathcal{E}=[E_1,\ldots,E_M]$, where $E_m$ is the probability that a tree splits on exposure $m$ and $\kappa$ is a hyperprior that controls the sparsity of exposures. This prior is motivated by \citet{Linero2018} but differs in that we select an exposure for the entire tree, while the former selects a variable to split on at a particular node of a tree. Details on setting $\kappa$ and a Bayes factor approach for exposure selection are give in Web Appendix C.
    
    Both $\mathcal{T}_{ai}$ and $S_{ai}$ are updated via Markov chain Monte Carlo (MCMC). New proposals for the structure of each tree in a tree-pair are the same as in TDLM. We also introduce a new proposal mechanism that switches the exposure, $S_{ai}$, considered by a tree. For each tree, we select with equal probability one of four different proposals: grow, prune, change, and switch-exposure. When switch-exposure is the update step we propose a new exposure $S_{ai}'$ from (\ref{eq:exposure-prob}). The decision to accept any of the four possible moves is made with the Metropolis-Hastings algorithm. The exposure, $S_{ai}$, can alternatively be updated with a Gibbs sampler, but at a high computational cost.

    The node-specific priors described in (\ref{eq:prior-tdlm}) shrinks tree-specific effects, but would apply the same variance parameters for all exposures. We introduce an alternative variance component pertaining specifically to each exposure or pair of two exposures. The prior on the node-specific effect, $\delta_{aib}$, at terminal node $\eta_{aib}$, which splits exposure $S_{ai}$, is defined as
    \begin{equation}\label{eq:dlmm-prior-marg}
        \delta_{ab}|\mu_{S_{ai}}^2,\nu^2,\sigma^2 
        \sim\mathcal{N}(0,\mu_{S_{ai}}^2\nu^2\sigma^2).
    \end{equation}
    Here, $\mu_{S_{ai}}$ is a local variance prior corresponding to the exposure, $S_{ai}$, being used by tree $\mathcal{T}_{ai}$. The prior for an interaction effect between terminal nodes $\eta_{a1b_1}$ and $\eta_{a2b_2}$ that split exposures $S_{a1}$ and $S_{a2}$, respectively, is
    \begin{equation}\label{eq:dlmm-prior-interact}
        \delta_{ab_1b_2}|\mu_{S_{a1}S_{a2}}^2,\nu^2,\sigma^2 
        \sim\mathcal{N}(0,\mu_{S_{a1}S_{a2}}^2\nu^2\sigma^2).
    \end{equation}
    Here, $\mu_{m}\sim\mathcal{C}^+(0,1)$ and $\mu_{m_1m_2}\sim\mathcal{C}^+(0,1)$ represent the variance components for the main effects of exposure $m$ or the interaction of exposures $m_1$ and $m_2$, respectively.  Because the same variance components are used in every tree or tree-pair that estimates the same exposure main or interaction effects, it allows the model to separate shrinkage on the main and interaction effects. Priors on $\boldsymbol\gamma,\nu,$ and $\sigma$ are the same as in TDLM. Full details of the MCMC algorithm used to sample from the posterior distribution of TDLMM are given in Web Appendix B.
    
    Selection and shrinkage are complementary in TDLMM. Shrinkage can apply to either the interaction effect, main effect, or both. Selection removes exposures from the model entirely.

    \subsection{Marginal DLM effects with TDLMM}\label{sec:post_dlm}
    Due to the interactions between exposures in TDLMM, the effects of each exposure are dependent on the levels of the other exposures. To estimate the main effect of each exposure $m$ on the outcome while accounting for co-exposures, we marginalize the DLMM at specified levels of all exposures. Fixing the levels of all exposures to be $\widetilde{x}_1,\ldots,\widetilde{x}_M$, the marginal effect of exposure $m$ at time $t$ is defined as
    \begin{equation}\label{eq:marg_exp_calc}
        \widetilde{\theta}_{mt}(\widetilde{x}_1,\ldots,\widetilde{x}_M)=
        \theta_{mt} + 
        \sum_{m'=1}^{m} \sum_{t'=1}^T
        \widetilde{x}_{m'}\theta_{m'mt't}
        +\sum_{m'=m}^M \sum_{t'=1}^T
        \widetilde{x}_{m'}\theta_{mm'tt'}.
    \end{equation}
    Marginalization by integrating out other exposure effects amounts to evaluating (\ref{eq:marg_exp_calc}) at the empirical mean of each exposure. Because the within-exposure interaction represents a possible nonlinear effect, this can cause interpretation problems when calculating $\widetilde{\theta}_{mt}$ (a linear effect). In this case, estimating a contrast, such as effect due to an interquartile range (IQR) change in exposure, is a more reasonable approach. 
    
    \subsection{Logistic regression}
    Binary outcomes are frequently used in environmental epidemiology. We propose a logistic regression method for TDLM and TDLMM. Details are available in Web Appendix B.5.

    \subsection{Prior specification}\label{sec:prior-setup}
    For tree structure parameters we follow \citet{Chipman2012} setting $\alpha=0.95$ and $\beta=2$. Altering these parameters did not improve performance. For TDLM we set $A=20$ trees. For TDLMM we set $A=20$ and $\kappa=1.089$ to achieve prior inclusion probability for exposure $m$ of 0.9. Additional details are provided in Web Appendix C.

    \section{Simulation}
    We considered two simulation scenarios. The first validated TDLM and compared to established DLM methods. The second evaluated TDLMM for a mixture of five exposures. In Web Appendix D we provide additional simulations to justify shrinkage components of our models. All simulations can be reproduced with \texttt{R} package \texttt{dlmtree}.

    \subsection{Single exposure and binary outcome}\label{sec:sim-TDLM}
    We first considered a binary outcome with a single exposure (PM$_{2.5})$ generated by
    \begin{equation}
        \label{eq:sim-scen1}
        y_i|p_i\sim\mbox{Bernoulli}(p_i),\quad
        \mbox{logit}(p_i)=c_1+0.1\cdot[f_1(\mathbf{x}_i)+\mathbf{z}_i^T\boldsymbol{\gamma}],
    \end{equation}
    with simulated DLM effect $f_1(\mathbf{x}_i)=\sum_{t=s}^{s+7} x_{it}$. This defines a DLM such that $\theta_t=0.1$ from times $s$ to $s+7$ and zero otherwise. Starting time $s$ was drawn uniformly from $\{1,\ldots,T-7\}$. We set $c_1$ for each simulation replicate such that the mean of $p_i$, denoted $\overline{p}$, was approximately $0.5$ or $0.05$. A binary outcome is used here as one of the comparison methods was only available for a single exposure in a logistic model. In Web Appendix D.3 we replicated this simulation using a smooth DLM effect.
    
    We simulated 100 data sets of sample size $n=$ 5,000. For each data set we sampled exposure observations from the cohort described in Section \ref{sec:data} with lengths $T=37$, and centered and scaled all exposure data. We also generated ten covariates  (five  standard  normal,  five  binomial  with  probability 0.5) and corresponding regression coefficients from standard  normal.
    
    In this scenario, we compared single exposure models TDLM to penalized spline DLM \citep{Gasparrini2017} with cubic regression splines and the critical window variable selection (CWVS) method \citep{Warren2020CriticalBirth}. In addition, we compared to TDLMM with all 5 exposures from our data analysis. For TDLMM, we included time-resolved measurements of four additional exposures that were not included in any other model and have no effect on the outcome in this design. Using the multi-exposure model in a single exposure setting explored the loss of efficiency that results when only one of five correlated predictors affects the outcome. The results for CWVS are based on the DLM estimation when a critical window is identified (DLM$|$cw) and when the probability of a critical window is greater than 0.5 (Pr(cw)$>$.5), as described by \citet{Warren2020CriticalBirth}. MCMC chains were run for 10,000 iterations after a burn-in period of 5,000 and thinned to every 5th iteration.
    
    The objective of our method is to estimate the distributed lag effect and identify critical windows. We therefore focus our simulation on estimation of these quantities rather than predictive performance, which is the focus of many BART models.
    
    Results from scenario one are given in Table \ref{tab:sim_scen_1}. We calculated the marginal DLM root mean square error (RMSE) $=\sqrt{\sum_{t=1}^{37}(\widetilde{\theta}_t-\widehat{\widetilde{\theta}}_t)^2\big/37}$ and coverage based on the estimated marginal distributed lag effect, $\widehat{\widetilde{\theta}}_t$. For TDLMM we calculated RMSE and coverage of PM$_{2.5}$ only. We also evaluated the probability that a model detects a true critical window (TP) or places a critical window where the true effect is zero (FP). For TDLMM, FP considered the marginal effects of all exposures. Finally, we evaluated precision of critical window identification, calculated $\mbox{TP}/(\mbox{TP}+\mbox{FP})$. 
    
    \begin{table}[!ht]
        \centering
        \caption{Results for simulation scenario one: binary outcome with single exposure DLM effect. We compare Gaussian process (CWVS) and penalized cubic regression spline (DLMcr) DLMs with our treed DLM and DLMM methods under conditions of a frequent ($\overline{p}=0.5$) or infrequent ($\overline{p}=0.05$) binary outcome. The first two columns describe DLM estimation, which refers to estimation of the marginal exposure effects, $\widetilde{\theta}_t$, for the active exposure. The final three columns describe critical window identification, which is summarized by the probability DLM 95\% credible intervals do not contain zero at correct (TP) or incorrect (FP) time periods, as well as $\mbox{Precision}=\mbox{TP}/(\mbox{TP}+\mbox{FP})$.}
        \begin{tabular}{@{}rcrrcrrr@{}}
            \toprule[2pt]
            &&
            \multicolumn{2}{c}{DLM Estimation}&\phantom{}&
            \multicolumn{3}{c}{Window Identification}\\
            \cmidrule{3-4} \cmidrule{6-8}
            Model&& RMSE$\;\times100$ & Coverage & & \phantom{}TP & \phantom{}FP & Precision \\ 
            \hline 
            \multicolumn{8}{l}{$\overline{p}=0.5$}\\
            CWVS: dlm$|$cw &  & 2.10 & 0.97 &  & 0.99 & 0.03 & 0.97 \\ 
            CWVS: p$>$0.5 &  & 2.10 & 0.97 &  & 0.99 & 0.07 & 0.93 \\ 
            DLMcr &  & 1.83 & 0.77 &  & 1.00 & 0.17 & 0.85 \\ 
            TDLM &  & 1.22 & 0.98 &  & 0.98 & 0.01 & 0.99 \\ 
            TDLMMadd &  & 1.31 & 0.97 &  & 0.99 & 0.02 & 0.98 \\ 
            TDLMMns &  & 1.47 & 0.96 &  & 1.00 & 0.03 & 0.97 \\ 
            TDLMM &  & 1.39 & 0.96 &  & 0.99 & 0.03 & 0.97 \\ 
            \multicolumn{8}{l}{$\overline{p}=0.05$}\\
            CWVS: dlm$|$cw &  & 3.36 & 0.99 &  & 0.72 & 0.02 & 0.97 \\ 
            CWVS: p$>$0.5 &  & 3.36 & 0.99 &  & 0.94 & 0.11 & 0.90 \\ 
            DLMcr &  & 2.64 & 0.72 &  & 1.00 & 0.21 & 0.83 \\ 
            TDLM &  & 2.19 & 0.96 &  & 0.88 & 0.02 & 0.98 \\ 
            TDLMMadd &  & 2.39 & 0.93 &  & 0.91 & 0.03 & 0.97 \\ 
            TDLMMns &  & 2.49 & 0.93 &  & 0.88 & 0.04 & 0.96 \\ 
            TDLMM &  & 2.59 & 0.91 &  & 0.83 & 0.04 & 0.95 \\ 
            \bottomrule[2pt]
            \bigskip
        \end{tabular}
        \label{tab:sim_scen_1}
    \end{table}
    
    The tree-based methods were the most accurate in estimating the distributed lag function; they had lowest RMSE and maintained near 95\% coverage. The penalized DLM had low RMSE, but poor coverage due to the wiggliness of splines, which also resulted in misclassification of critical windows. The CWVS model maintained high coverage of the distributed lag function, but was not as accurate in estimating the effects as evidenced by RMSE.
    
    In both $\overline{p}$ settings, the mixture models outperformed the single exposure spline-based DLM and CWVS models in terms of RMSE while maintaining near nominal coverage. These are important findings because the cost of including additional exposures into the treed models is minimal in terms of distributed lag function estimation. In TDLMMns when $\overline{p}=0.5$ the posterior inclusion probability of PM$_{2.5}$ was 1. This decreased slightly ($0.89$) when $\overline{p}=0.05$. For other exposures, the posterior inclusion probability was $<0.01$ in both $\overline{p}$ settings.
    
    Comparing models in terms of critical window detection, TDLM, TDLMM, and CWVS: dlm$|$cw were the most precise models at identifying critical windows. In the more difficult $\overline{p}=0.05$ setting, the increased complexity of TDLMM was associated with a slight decrease in TP, but the method retained high precision due to the very low FP rate. CWVS: $p>0.5$ had higher TP at the cost of higher FP. Spline methods were less precise with higher FP.
    
    Results from the supplemental simulation with a smooth distributed lag function were similar and are presented in Web Appendix D.3. In both the current and smooth simulations, TDLM and TDLMM had lower RMSE, closer to nominal coverage, and higher precision in identifying critical windows compared to the alternative estimators.
    
    \subsection{Multiple exposures and a continuous outcome}
    We considered a continuous response with an exposure main effect from PM$_{2.5}$ and an interaction effect between PM$_{2.5}$ and NO$_2$. The outcome was generated as
    \begin{equation}
        \label{eq:sim-scen2}
        y_i=c_2 f_2(\mathbf{x}_{i1},\mathbf{x}_{i2})+\mathbf{z}_i^T\boldsymbol{\gamma}+\epsilon_i
    \end{equation}
    where $c_2$ is a scalar such that the variance of $c_2 f_2$ equals one. The simulated effect was
    \begin{equation}
        f_2(\mathbf{x}_{i1},\mathbf{x}_{i2})= \sum_{t=s_1}^{s_1+7}x_{i1t}+ 0.025\sum_{t_1=s_1}^{s_1+7} \sum_{t_2=s_2}^{s_2+7}x_{i1t_1}x_{i2t_2}
    \end{equation}
    for starting times $s_1$ and $s_2$ each drawn uniformly from $\{1,\ldots,T-7\}$. This scenario consists of a main effect with a critical window of length eight for the PM$_{2.5}$ exposure and an interaction effect between exposure to PM$_{2.5}$ at times $s_1$ to $s_1+7$ and NO$_2$ at times $s_2$ to $s_2+7$. We drew $\epsilon_i$ independently for each observation from $\mathcal{N}(0,\sigma^2)$ such that $\sigma^2\in\{25, 50, 100\}$ is the noise-to-signal ratio. In this scenario, we compared TDLMM, TDLMMns (no-self interactions) and TDLMMadd (additive DLMs) using all exposure measurements from our data analysis. No other methods are currently available that would offer a direct comparison.
    
    Table \ref{tab:sim_scen_2} compares the marginal DLM RMSE and coverage for active exposures PM$_{2.5}$ and NO$_2$. As in scenario one, we describe TP and FP for critical window detection. We found that all variants of TDLMM had low RMSE for estimating the marginalized effects of PM$_{2.5}$ and NO$_2$. The additive model had below nominal coverage for the marginal effect of NO$_2$ due to lack of appropriate interaction terms. We found that TDLMMns had higher power than full TDLMM. Additive TDLMM had the highest TP rate, but increased FP for NO$_2$. All TDLMM variants show zero FP for nonactive exposures. Overall, the tree-based DLMMs had high precision for identifying critical windows.
    
    \begin{table}[!ht]
        \centering
        \caption{Results for simulation scenario two: main effect of PM$_{2.5}$ with PM$_{2.5}-$NO$_2$ interaction. The first four columns describe RMSE and coverage of the estimated marginal distributed lag effects, $\widetilde{\theta}_t$, for PM$_{2.5}$ and NO$_2$ compared to the true simulated marginal effects. The final five columns compare critical window detection for marginal effects of PM$_{2.5}$ and NO$_2$, which are summarized by the probability that the 95\% CI does not include zero at a correct (TP) or incorrect (FP) time period. `Other' corresponds to the three additional exposure not related to the outcome.}
        \begin{tabular}{@{}rcrrcrrcrrcrrr@{}}
            \toprule[2pt]
            &\phantom{}&
            \multicolumn{2}{c}{RMSE$\;\times100$}&\phantom{}&
            \multicolumn{2}{c}{Coverage}&\phantom{}&
            \multicolumn{2}{c}{TP}&\phantom{}&
            \multicolumn{3}{c}{FP}\\
            \cmidrule{3-4} \cmidrule{6-7} \cmidrule{9-10} \cmidrule{12-14}
            Model&& PM & NO$_2$ && PM & NO$_2$ && PM & NO$_2$ && PM & NO$_2$&Other \\ 
            \hline
            \multicolumn{14}{l}{$\sigma^2=25$}\\
            TDLMMadd &  & 3.59 & 4.19 &  & 0.95 & 0.84 &  & 0.94 & 0.82 &  & 0.03 & 0.07 & 0.00 \\ 
            TDLMMns &  & 3.57 & 4.27 &  & 0.97 & 0.96 &  & 0.91 & 0.63 &  & 0.02 & 0.03 & 0.00 \\ 
            TDLMM &  & 3.55 & 4.33 &  & 0.98 & 0.97 &  & 0.87 & 0.51 &  & 0.01 & 0.02 & 0.00 \\ 
            \multicolumn{14}{l}{$\sigma^2=50$}\\
            TDLMMadd &  & 4.62 & 4.64 &  & 0.92 & 0.83 &  & 0.86 & 0.52 &  & 0.04 & 0.05 & 0.00 \\ 
            TDLMMns &  & 4.61 & 4.75 &  & 0.96 & 0.97 &  & 0.76 & 0.31 &  & 0.02 & 0.02 & 0.00 \\ 
            TDLMM &  & 4.50 & 4.79 &  & 0.97 & 0.97 &  & 0.62 & 0.22 &  & 0.01 & 0.01 & 0.00 \\ 
            \multicolumn{14}{l}{$\sigma^2=100$}\\
            TDLMMadd &  & 5.75 & 5.07 &  & 0.89 & 0.84 &  & 0.47 & 0.22 &  & 0.03 & 0.03 & 0.00 \\ 
            TDLMMns &  & 5.85 & 5.21 &  & 0.93 & 0.95 &  & 0.29 & 0.10 &  & 0.02 & 0.01 & 0.00 \\ 
            TDLMM &  & 5.73 & 5.21 &  & 0.94 & 0.96 &  & 0.24 & 0.05 &  & 0.01 & 0.01 & 0.00 \\ 
            \bottomrule[2pt]
            \multicolumn{7}{l}{$^*\mbox{RMSE}\times100$}
            \bigskip
        \end{tabular}
        \label{tab:sim_scen_2}
    \end{table}
    
    Table \ref{tab:sim_scen_2b} summarizes exposure and interaction posterior inclusion probabilities averaged across simulation replicates. For the correct exposures, all variants of TDLMM had posterior inclusion probability above the prior inclusion probability (0.9) in low and medium noise settings. For nonactive exposures, the posterior inclusion probability was below that of the correct exposures. The differences were less pronounced in TDLMMadd. Posterior inclusion probabilities for the correct interaction were consistently higher than for other interactions.
    
    \begin{table}[!ht]
        \centering
        \caption{Exposure and interaction posterior inclusion probabilities. The values in this table describe the probability at least one tree or tree-pair in the ensemble estimates the effects for a specific exposure or interaction, respectively. Nonactive exposures or interactions are summarized together as `Other'.}
        \begin{tabular}{@{}rcrrrcrr@{}}
            \toprule[2pt]
            &\phantom{ab}&
            \multicolumn{3}{c}{Main Effect}&\phantom{ab}&
            \multicolumn{2}{c}{Interaction}\\
            \cmidrule{3-5} \cmidrule{7-8}
            Model&& \phantom{a}PM &\phantom{a}NO$_2$ & Other && PM$-$NO$_2$ & Other\\ 
            \hline
            \multicolumn{8}{l}{$\sigma^2=25$}\\
            TDLMMadd &  & 1.00 & 1.00 & 0.78 &  & - & - \\ 
            TDLMMns &  & 1.00 & 1.00 & 0.17 &  & 0.94 & 0.06 \\ 
            TDLMM &  & 1.00 & 0.99 & 0.39 &  & 0.85 & 0.22 \\ 
            \multicolumn{8}{l}{$\sigma^2=50$}\\
            TDLMMadd &  & 1.00 & 0.99 & 0.82 &  & - & - \\ 
            TDLMMns &  & 0.99 & 0.95 & 0.26 &  & 0.83 & 0.10 \\ 
            TDLMM &  & 0.99 & 0.96 & 0.43 &  & 0.80 & 0.23 \\  
            \multicolumn{8}{l}{$\sigma^2=100$}\\
            TDLMMadd &  & 0.98 & 0.96 & 0.85 &  & - & - \\ 
            TDLMMns &  & 0.88 & 0.81 & 0.38 &  & 0.61 & 0.15 \\ 
            TDLMM &  & 0.89 & 0.85 & 0.52 &  & 0.58 & 0.25 \\  
            \bottomrule[2pt]
            \bigskip
        \end{tabular}
        \label{tab:sim_scen_2b}
    \end{table}

    \section{Analysis of Colorado Birth Cohort Data}
    We applied TDLMM to BWGAZ for our Denver, CO metro area dataset. The analysis included five exposures and covariates described in Section \ref{sec:data} with $T=37$ weeks of exposure measurements corresponding to the first 37 weeks of gestation. In addition, our analysis controlled for year and month of conception, census tract elevation, and a county-specific intercept. TDLMM used the prior specification described in Section \ref{sec:prior-setup}. The model ran for 100,000 iterations after 10,000 burn-in and was thinned to every 5th iteration. For this analysis, we used TDLMMns for ease of interpretation of the linear effects. We compared to TDLMM and found few differences (see Web Appendix E.2).

    \subsection{Exposure and interaction selection}
    Four exposures had posterior inclusion probabilities above the prior inclusion probability (0.9): PM$_{2.5}$ ($>0.99$), SO$_2$ ($>0.99$), CO (0.989), and temperature ($>0.99$). NO$_2$ was included into the model at a much lower rate (0.428). The posterior inclusion probability for the PM$_{2.5}-$temperature interaction was 0.988. The next highest posterior inclusion probabilities for interactions were CO$-$temperature (0.69) and SO$_2-$CO (0.6). The seven remaining interactions had posterior inclusion probabilities below 0.5.

    \subsection{Marginal exposure associations with BWGAZ}\label{sec:marg_data}
\begin{figure}[!ht]
        \centering
        \includegraphics[width=\textwidth]{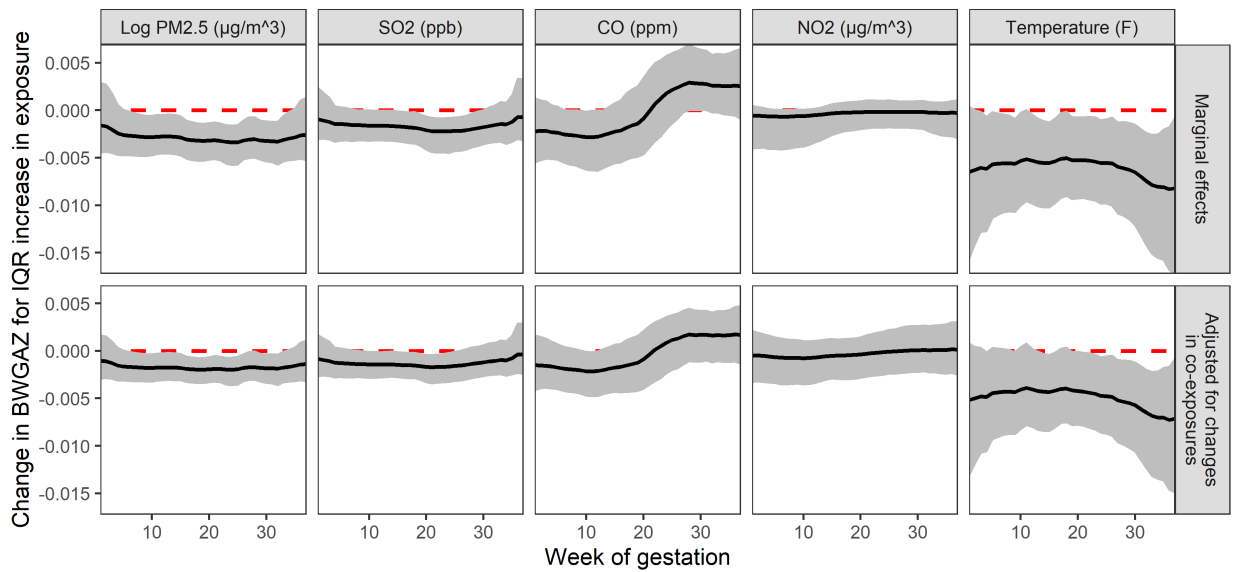}
        \caption{Posterior mean distributed lag function (black line) for each exposure (columns) with 95\% credible interval (grey area) of the effect. The top row shows the estimated marginal effect for an IQR increase in exposure, holding other exposures at their empirical mean. The bottom row shows the estimated change in BWGAZ for a first to third quartile change in one exposure along with the expected changes in all other exposures due to correlation with the exposure of interest.}
        \label{fig:bwgaz_marg_effects}
    \end{figure}

    The top row of Figure \ref{fig:bwgaz_marg_effects} shows the estimated marginal distributed lag function of each exposure with other exposures fixed to their empirical means. The results identify critical windows to PM$_{2.5}$ during weeks $6-34$, SO$_{2}$ during weeks $10-29$, CO during weeks $9-11$ and temperature across the entire pregnancy. The critical windows show that increased exposure is associated with decreased BWGAZ.
    
\begin{figure}[!ht]
        \begin{center}
            \subfigure[PM$_{2.5}-$temperature interaction surface.]{
                \label{fig:bwgaz_pm_temp_int}
                \includegraphics[width=.47\textwidth]{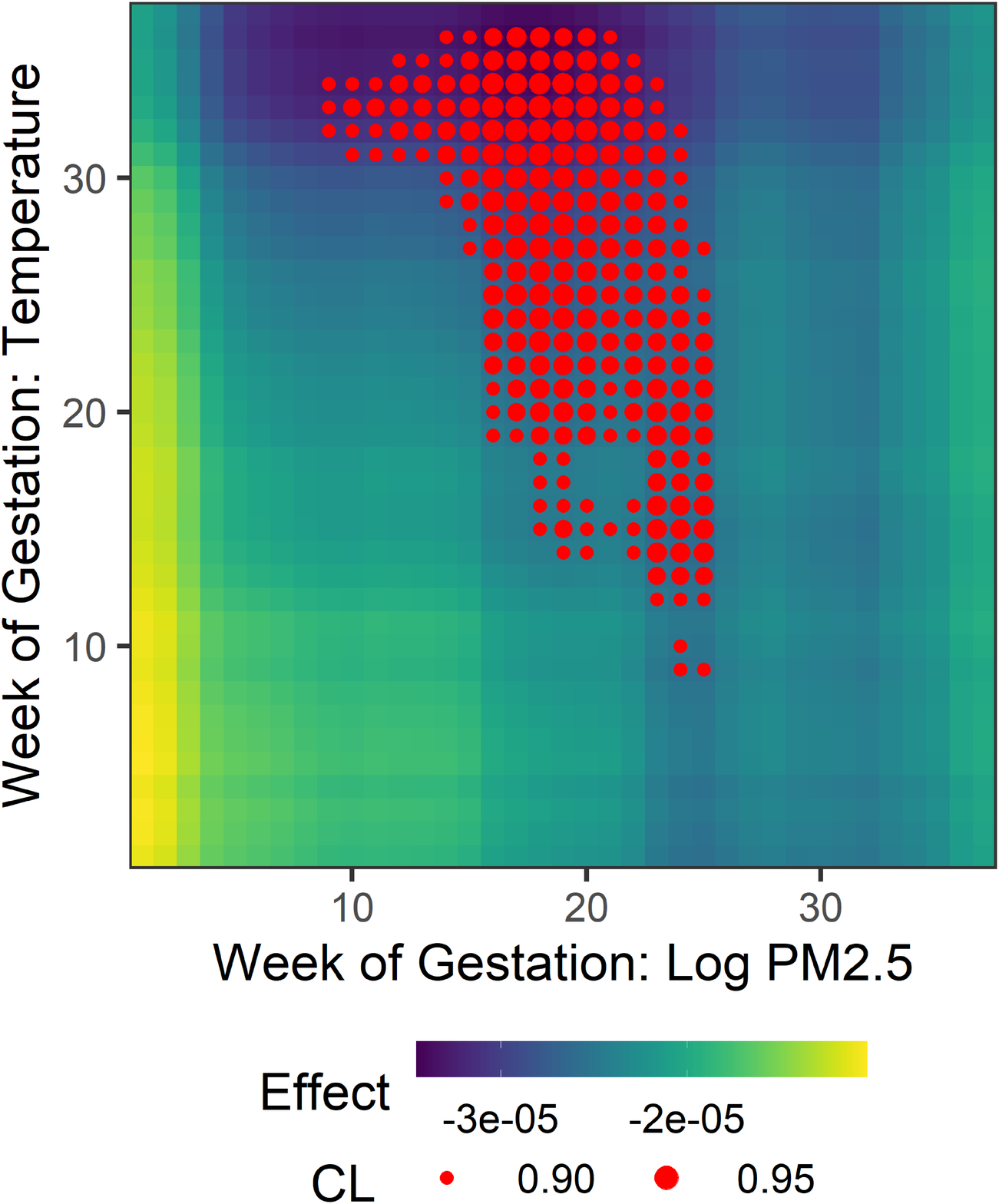}
            }\quad
            \subfigure[Marginal DLMs showing interaction.]{
                \label{fig:bwgaz_marg_int}
                \includegraphics[width=.47\textwidth]{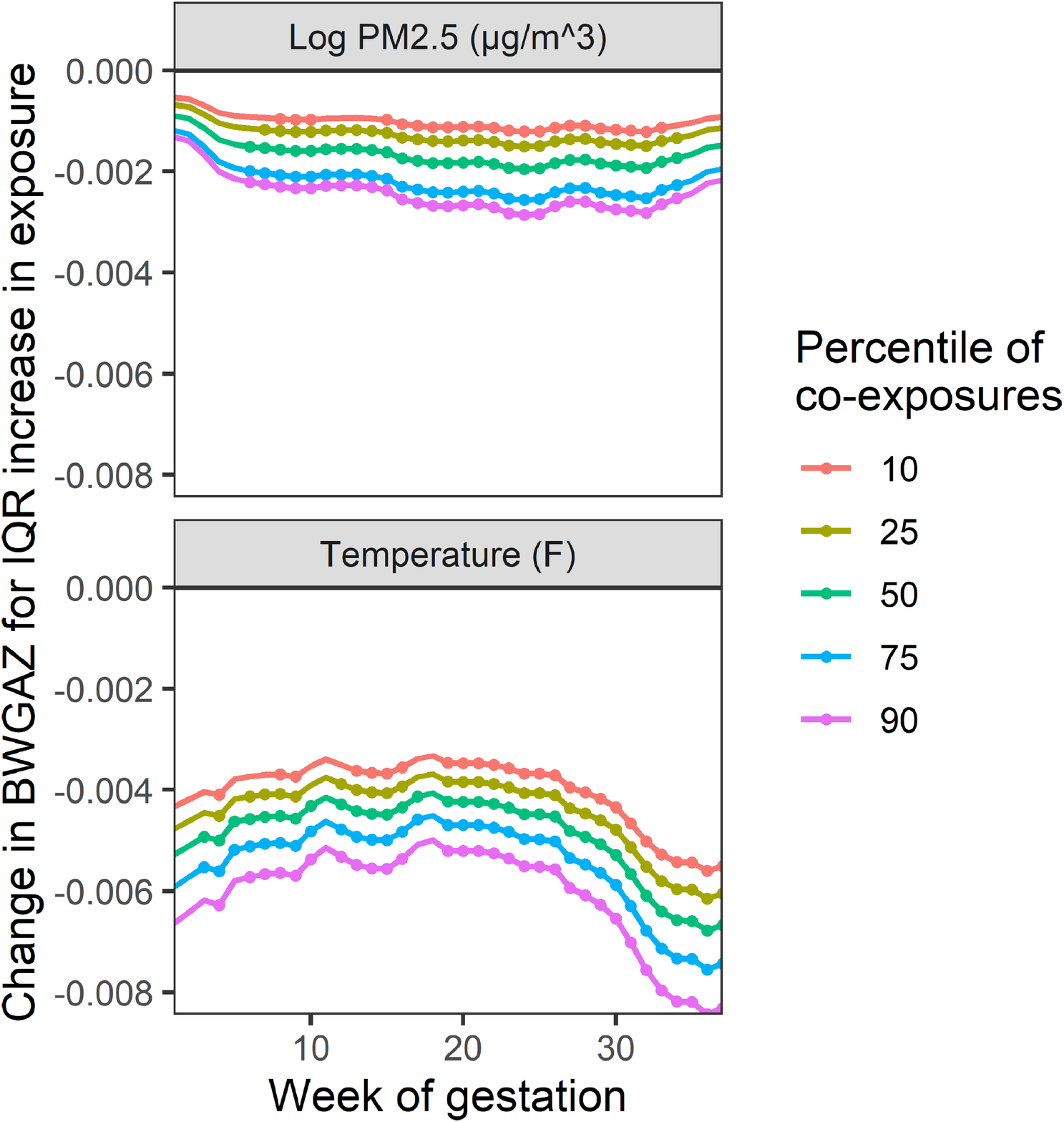}
            }
            \caption{Panel (a) shows the estimated PM$_{2.5}-$temperature interaction effects.  The color of each cell indicates the sign and direction of the interaction effect between PM$_{2.5}$ at one time and temperature at another time. The points indicate estimated interactions where the credible interval does not contain zero where larger points correspond to higher probability credible intervals. Panel (b) shows the estimated marginal distributed lag function for an IQR increase in the indicated exposure, when all other exposures are fixed at a given percentile (color of lines) of their empirical distributions. Points on each line indicate where the 95\% credible interval of the marginal effect does not include zero.}
        \end{center}
    \end{figure}

    TDLMM allows us to estimate interactions between time-resolved predictors. Figure \ref{fig:bwgaz_pm_temp_int} shows the presence of an interaction between log PM$_{2.5}$ and temperature. This interaction indicates that elevated exposure to PM$_{2.5}$ during weeks $15-25$ results in an increase in the magnitude of the temperature effect during weeks $19-35$. These interactions occur at the same times (e.g. weeks $19-25$) as well as different times (e.g. log PM$_{2.5}$ in week 20 and temperature in week 30) and may be explained by the priming hypothesis. Figure \ref{fig:bwgaz_marg_int} shows the marginal distributed lag functions for PM$_{2.5}$ and temperature at varying percentiles of co-exposures. We see a larger effect of PM$_{2.5}$ and changes in critical windows when temperature and other exposures are fixed at a higher percentile. The effect of temperature also shows changes in the magnitude and timing of critical windows at increased levels of other exposures.  Additional interaction plots are included in Web Appendix E.1 and show no evidence of meaningful interactions.
    
    \subsection{Accounting for changes in co-exposures}
    Due to high correlation among exposures in our analysis (ranging from $-0.55$ to $0.69$ at the same week), a change in any one exposure will likely be accompanied by simultaneous changes in co-exposures. This suggests that the marginal results in Section \ref{sec:marg_data}, which assumes that co-exposures are held at their empirical means, should only be interpreted in a narrow range. As an alternative, we estimate distributed lag functions that account for simultaneous changes in co-exposures. This provides another way to look at the results from the same model.
    
    
    Let $x_{m(q)}$ represent the q$^{th}$ quantile of the empirical distribution of exposure $m$. The expected change in the outcome for an IQR change in exposure $m$ at time $t$, conditional on the expected co-occurring changes in all other exposures can be written
    \begin{multline}\label{eq:exp_dlm}
        \mbox{E}\left[Y \Big|
        \widetilde{\mathbf{X}}_t = 
        \mbox{E}\left\{\mathbf{X}_t \big|
        x_{mt}=x_{m(0.75)}\right\}, \widetilde{\mathbf{X}}_{[t]}=\overline{\mathbf{x}}, \mathbf{z}=\mathbf{z}_0\right]\\
        -\mbox{E}\left[Y \Big|
        \widetilde{\mathbf{X}}_t = 
        \mbox{E}\left\{\mathbf{X}_t \big|
        x_{mt}=x_{m(0.25)}\right\}, \widetilde{\mathbf{X}}_{[t]}=\overline{\mathbf{x}}, \mathbf{z}=\mathbf{z}_0\right].
    \end{multline}
    Here, $\widetilde{\mathbf{X}}_t=\{\widetilde{x}_{1t},\ldots,\widetilde{x}_{Mt}\}$ defines the values of all exposures at time $t$ while $\widetilde{\mathbf{X}}_{[t]}$ is the collection of exposure measurements at all time points except $t$. We set each element of $\widetilde{\mathbf{X}}_{[t]}$ equal to the empirical mean for that exposure ($\overline{\mathbf{x}}$) to isolate the effect only at the time of interest. The value of $\mathbf{z}_0$ is set at a population baseline, but does not influence the expected change because it does not interact with exposure measurements. To estimate the values of co-exposures at time $t$, given by $\mbox{E}[\mathbf{X}_t \big|x_{mt}=x_{m(q)}]$, we fit penalized spline models (cubic splines with 5 degrees of freedom) between pairs of exposures, estimated using restricted maximum likelihood as described by \citet{Wood2017GeneralizedR}. Specifically, we considered measurements for exposure $m$ as the only predictor and fit separate models for each co-exposure (e.g. log PM$_{2.5}$ was used as the predictor for NO$_2$, SO$_2$, CO, and temperature in four separate models). This model matches our assumption of a simple, smooth relationship between pairs of exposures. Using the model fits, we predicted all co-exposures at the 25$^{th}$ and 75$^{th}$ percentiles of exposure $m$. This process was repeated for every $m\in\{1,\ldots,5\}$.
    
    The results of this analysis adjusting for changes in co-exposures are shown in the bottom row of Figure \ref{fig:bwgaz_marg_effects}. This can be interpreted as the expected change in BWGAZ associated with an IQR change in one pollutant and the expected change in the four other co-exposures at the same time point. There are several important takeaways. First, the shape of estimated distributed lag functions are similar. Second, the confidence intervals take into account the uncertainty in the estimated main effects and interactions of co-exposures. That uncertainty is not included in the estimates presented in the top row of Figure \ref{fig:bwgaz_marg_effects} because the level of co-exposures remain fixed. Third, the associations between BWGAZ and exposure to PM$_{2.5}$, SO$_2$ and temperature are persistent after accounting for expected changes in co-exposures.
    
    A change in PM$_{2.5}$ from the 25$^{th}$ to 75$^{th}$ percentile of exposure is an increase from 6.12 to 8.67 $\mu g/m^3$. Considering the association between PM$_{2.5}$ and BWGAZ after adjusting for co-exposures finds a critical window during weeks $6-33$, which is similar to the critical window found in the marginal distributed lag function. The cumulative effect of PM$_{2.5}$, or the effect of a simultaneous IQR increase in every week of pregnancy, adjusting for co-exposures is associated with an average change in BWGAZ of $-0.064$ (95\% CI: $[-0.094, -0.035]$). The cumulative effect of an IQR increase in SO$_2$ (0.94 to 1.90 ppb) is associated with a change in BWGAZ of $-0.049$ (95\% CI: $[-0.076, -0.019]$), with a critical window during weeks $9-25$.
    
    
    Our findings of an inverse relationship between PM$_{2.5}$ and birth weight are consistent with a meta-analysis done by \citet{Stieb2012AmbientMeta-analysis}. Similar findings exist for a relationship between SO$_2$ and birth weight \citep{Dugandzic2006TheStudy}, but are less consistent \citep{Stieb2012AmbientMeta-analysis}. We note that the majority of comparison studies do not account for effects of co-occurring exposures or interactions over time and no previous study has considered time-resolved measures of five exposures including interactions.

    \section{Discussion}
    In this work we propose TDLMM to estimate the association between mixtures of environmental exposures observed at high-temporal resolution on birth and children’s health outcomes. TDLMM, and the reduced version TDLM, have several innovative statistical features. The models introduce structured regression trees that estimate a constrained DLM. The mixtures approach, TDLMM, uses an ensemble of tree pairs that parameterize both main effects and pairwise interactions between time-resolved predictors.
    
    In a simulation study we demonstrated that our single exposure tree-based DLM outperformed established methods for estimating the DLM in terms of estimation of the exposure-response function and precision in identifying critical windows. Moreover, our tree-based mixture approach, TDLMM, that included additional exposures not associated with the outcome also outperformed the established single exposure methods in the single exposure setting. This is in large part due to the methods' ability to select out or shrink the effects of exposures that are not associated with the outcome. Using real exposure data, TDLMM identifies critical windows, selects the proper exposures, and estimates the exposure-response function in a simulation scenario with five time-resolved predictors.
    
    We applied TDLMM to analyze associations between Denver, CO area birth weight and five environmental exposures experienced weekly by mothers during gestation. This data set included 195,701 full term births with estimated conception dates from 2007 through 2015. This data analysis to estimate the main and interaction effects due to a mixture of five environmental exposures observed weekly during the first 37 weeks of gestation produced several key takeaways. We identified PM$_{2.5}$, SO$_2$ and temperature effects based on the effect of one exposure conditional on fixed co-exposures and based on a change in one exposure accounting for simultaneous changes in co-exposures. In addition, we found substantial evidence of an interaction between PM$_{2.5}$ and temperature. Independently, birth weight changes due to exposure appear small, but when combined with other risk factors (including those often correlated with increased pollution) may have a substantial impact on birth weight, which may increase the prevalence of associated comorbidities. Furthermore, there is evidence that the effects of air pollution are larger at the lower quantiles of birth weight---arguably the more susceptible population \citep{Lamichhane2020QuantileWeight}.
    
    Our analysis of five pollutants observed weekly throughout pregnancy and birth weight is the first analysis to identify critical windows to a mixture observed at high-temporal resolution within a distributed lag mixture model framework. Previous studies have either estimated critical windows for a mixture observed at high temporal resolution in an additive DLM setting \citep{Figueroa-Romero2020EarlySclerosis,Horton2018DentineBehavior} or to a mixture observed at one or a small number of time points \citep{Levin-Schwartz2019Time-varyingChildren,Chiu2018EvaluatingApproaches}.  Data driven methods that allow for precision in identifying critical windows due to environmental mixtures have the potential to open doors to discovery and understanding in biological science \citep{Wright2017EnvironmentHealth}. As the size and resolution of the exposure data available continues to grow, our method fills a critical research gap in statistical methods for epidemiology and environmental health in being able to estimate the effects of time-resolved measures of a mixture on a continuous or binary health outcome.

    \section*{Acknowledgements}
    This work was supported by National Institutes of Health grants ES029943 and ES028811. 
    
    These data were supplied by the Center for Health and Environmental Data Vital Statistics Program of the Colorado Department of Public Health and Environment, which specifically disclaims responsibility for any analyses, interpretations, or conclusions it has not provided.
    
    This work utilized resources from the University of Colorado Boulder Research Computing Group, which is supported by the National Science Foundation (awards ACI-1532235 and ACI-1532236), the University of Colorado Boulder, and Colorado State University.
    
    \bibliography{references.bib}
\end{document}